\documentclass[showkeys,preprint,showpacs,nopreprintnumbers,amsmath,amssymb,aps,prb]{revtex4}

\usepackage{graphicx}% Include figure files
\usepackage{bm}% bold math

\begin{document}
% \thispagestyle{myheadings}
% \markright{Submitted to Appl. Phys. Lett.}
% \markleft{Submitted to Appl. Phys. Lett.}

\pagenumbering{arabic}

\title{Large inverse tunneling magnetoresistance in
Co$_2$Cr$_{0.6}$Fe$_{0.4}$Al/MgO/CoFe magnetic tunnel junctions}

\author{A.~D. Rata}
\author{H. Braak}
\author{D.~E. B\"urgler}
\thanks{Corresponding author. Email: d.buergler@fz-juelich.de, phone:
+49 2461 614 214, FAX: +49 2461 614 443}
\author{C.~M. Schneider}
\affiliation{Institute of Solid State Research - Electronic
Properties (IFF-9) and \\ cni -- Center of Nanoelectronic Systems for
Information Technology, Research Center J\"ulich GmbH, D-52425
J\"ulich, Germany}

\date{\today}

\begin{abstract}
Magnetic tunnel junctions with the layer sequence
Co$_2$Cr$_{0.6}$Fe$_{0.4}$Al/MgO/CoFe were fabricated by magnetron
sputtering at room temperature (RT).  The samples exhibit a large
inverse tunneling magnetoresistance (TMR) effect of up to $-66$\%
at RT. The largest value of $-84$\% at 20\,K reflects a rather
weak influence of temperature. The dependence on the voltage drop
shows an unusual behavior with two almost symmetric peaks at
$\pm600$\,mV with large inverse TMR ratios and small positive
values around zero bias.
\end{abstract}


 \maketitle

%\newpage

The tunnel magnetoresistance effect (TMR), \textit{i.e.} the
resistance change upon application of a magnetic field in a
structure consisting of two ferromagnetic (FM) layers separated by
a thin insulating barrier, is subject of intense experimental
research due to its potential for spintronic
applications.\cite{wolf} The aim is to achieve a large TMR ratio
for practical output voltages in combination with a weak
temperature dependence. Recently, very high TMR ratios exceeding
200\% at room temperature (RT) have been reported for magnetic
tunnel junctions (MTJ) with fully epitaxial or highly oriented MgO
barriers and Fe,\cite{yuasa} CoFe,\cite{parkin} and CoFeB
\cite{djayaprawira} electrodes. Resonant tunneling and specific
features in the band structures of MgO and Fe-based alloys with
bcc structure give rise to a spin-filter effect, such that
predominantly highly spin-polarized states of the FM contribute to
the tunneling current.\cite{butler} Weak contributions from other
states and any breaking of the translational invariance parallel
to the interfaces (\textit{e.g.} interface roughness, defects)
limit the maximum achievable TMR ratio.\cite{zhang} Another
strategy for obtaining large TMR effects is to use FM electrodes
with an \textit{intrinsically} high spin polarization, such as
half-metallic ferromagnets featuring 100\% spin polarization of
the carriers.\cite{degroot} Among the vast family of the
half-metallic ferromagnets, the Heusler alloys are promising
candidates due to their high Curie temperatures well above
300~K.\cite{magprop}

In the last years, experimental efforts were concentrated on
improving the structure of Heusler thin films and the properties
of the interface between the Heusler electrode and the oxide
barrier. Relatively high TMR ratios have been obtained using
Co-based full-Heusler alloy thin films, \textit{e.g.} Co$_2$MnSi
\cite{hutten1,sakuraba1} and Co$_2$Cr$_{0.6}$Fe$_{0.4}$Al.
\cite{inomata,inomata1,marukame} Marukame \textit{et al.}
\cite{marukame2} prepared fully epitaxial
Co$_2$Cr$_{0.6}$Fe$_{0.4}$Al/MgO/CoFe MTJs with up to 90\% TMR at
RT. In this letter we report on large \textit{inverse} TMR values
obtained from Co$_2$Cr$_{0.6}$Fe$_{0.4}$Al/MgO/CoFe MTJ
structures.

Details on the preparation and characterization of
Co$_2$Cr$_{0.6}$Fe$_{0.4}$Al (CCFA) Heusler thin films have been
published elsewhere.\cite{rata} In contrast to previous reports
about the growth of CCFA films on MgO(001)
substrates,\cite{clemens1,clemens2,marukame} our CCFA films are
grown at RT and adopt a different crystalline orientation with
respect to the MgO(001) substrate.\cite{rata} The [011] direction
of the CCFA films is parallel to [001] of the MgO(001) substrate.
The CCFA films have the B2 structure, since the (111) reflection
is absent in the X-ray diffraction patterns. Magnetization
measurements of extended CCFA films reveal ferromagnetic ordering
with Curie temperatures up to 630\,K after annealing in vacuum at
773\,K. The total magnetic moment found is about $2.5\mu_{B}$ per
formula unit (f.u.). This value is small compared to the
theoretical bulk value of $3.8\mu_{B}$/f.u.,\cite{galanakis3} but
still comparable to the moments reported in
Refs.~\onlinecite{clemens1,clemens2} for films grown at elevated
temperature.

MTJs are prepared by magnetron sputtering at RT without breaking
the vacuum with the following layer sequence: MgO(100)/MgO(40
nm)/CCFA(25 nm)/MgO(3 nm)/CoFe(5 nm)/IrMn(15 nm).  A 40\,nm-thick
MgO seed layer is deposited on the MgO substrate to improve the
texture of the CCFA electrode.  CCFA and MgO are deposited by DC
and RF stimulated discharge, respectively, from stoichiometric
targets.  The sputtering rates monitored with a quartz balance are
calibrated by measuring the thicknesses of individual thin film
layers by X-ray reflectivity.  The completed stack is annealed
\textit{in-situ} for 1\,hour at 523\,~K in order to improve the
interface quality.  However, we cannot apply a magnetic field
during the \textit{in-situ} annealing.  Thus, the CoFe electrode
is not exchange-biased.  Junctions with an area from $3\times3$ up
to $15\times15$~$\mu$m$^{2}$ with crosse-bar electrodes are
patterned for magnetotransport measurements in the
current-perpendicular-plane (CPP) geometry by optical lithography.
The transport measurements are performed with a DC setup in the
standard 4-point geometry using a constant current source.  $I-V$
characteristics measured at RT in zero field show nonlinear,
\textit{i.e.} non-Ohmic behavior.

\begin{figure}
\includegraphics[width=7cm]{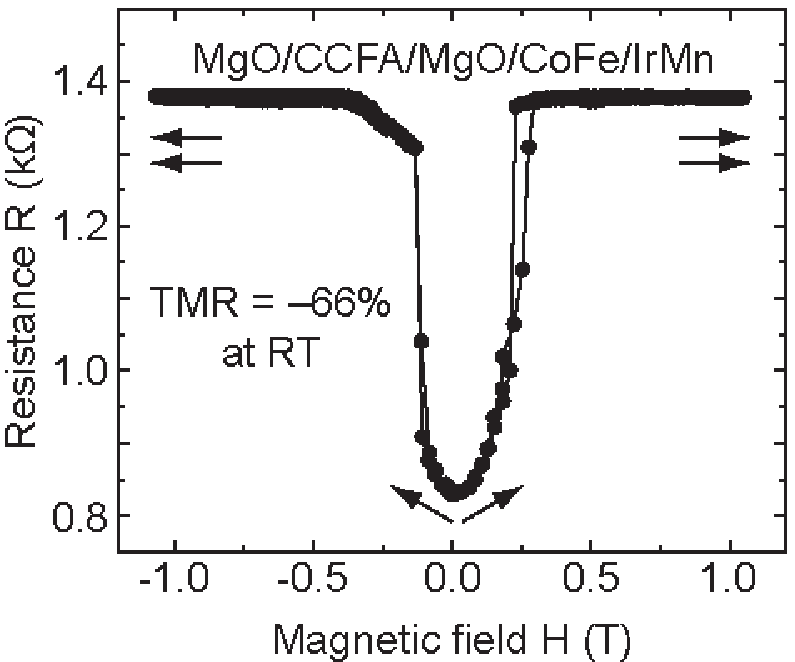}
\caption{Magnetoresistance curve of a $10\times 10$~$\mu$m$^{2}$
MTJ with layer sequence MgO/MgO(40 nm)/CCFA(25 nm)/MgO(3
nm)/CoFe(5 nm)/IrMn(15 nm).  The measurement is performed at RT
and yields an \textit{inverse} magnetoresistance of $-66$\% for
$\Delta V=+600$\,mV.} \label{f-mr}
\end{figure}

In Fig.~\ref{f-mr} we present a magnetoresistance curve measured
at RT on a CCFA/MgO/CoFe MTJ deposited on MgO(100) substrate with
a 40\,nm-thick MgO seed layer.  Obviously, we observe an
\textit{inverse} TMR effect.  The TMR ratio defined as in
Ref.~\onlinecite{detereza} as TMR = $(R_{AP}-R_{P})/R_{AP}$, where
$R_{AP}$ is the smallest resistance value in the antiparallel
magnetization configuration and $R_{P}$ denotes the highest
resistance in the saturated state, reaches $-66$\%.  The minimal
resistance times area product is about 80\,k$\Omega\mu$m$^{2}$.
The junction resistances are 1\,k$\Omega$ or larger in all samples
and, thus, clearly exceed the lead and contact resistances of
10-20\,$\Omega$. Therefore, there is no geometrical enhancement in
the TMR ratios. The TMR value obtained at RT is relatively large
for structures comprising a FM Heusler electrode.  The bell-shaped
MR curve depicted in Fig.~\ref{f-mr} suggests a noncollinear
orientation of the layers magnetizations for $H=0$ due to magnetic
coupling of the two FM layers (arrows in Fig.~\ref{f-mr}). This is
also confirmed by SQUID measurements, where independent switching
of the two electrodes is found to be hindered. Antiferromagnetic,
N\'{e}el-type and biquadratic coupling due to interface roughness
are the most likely coupling mechanisms.

\begin{figure}
\includegraphics[width=7cm]{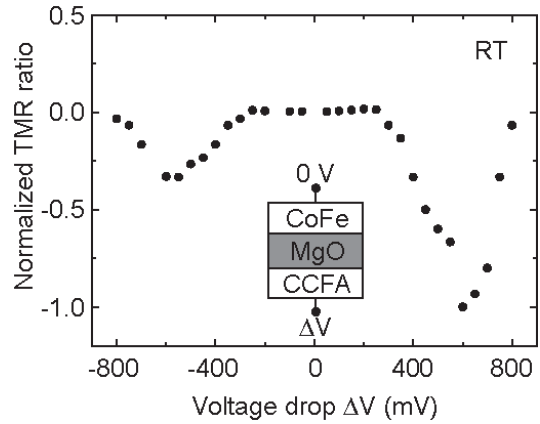}
\caption{Normalized TMR ratio of a MTJ with layer sequence
MgO/MgO(40 nm)/CCFA(25 nm)/MgO(3 nm)/CoFe(5 nm)/IrMn(15 nm) as a
function of the voltage drop $\Delta V$ across the junction
measured at RT. Inset: $\Delta V$ is defined with respect to the
CoFe electrode.} \label{f-voltage}
\end{figure}

Figure~\ref{f-voltage} shows the typical dependence of the TMR
ratio on the voltage drop $\Delta V$ across the MTJ measured in
the parallel configuration at $H=\pm1$\,T. $\Delta V$ is
experimentally controlled by varying the current bias supplied by
the constant current source and is defined with respect to the
CoFe electrode (see inset of Fig.~\ref{f-voltage}).  The inverse
TMR effect is dominant over a wide range of positive and negative
$\Delta V$.  Only for $|\Delta V|<300$\,mV we obtain a small
positive TMR ratio of less than 1\%. From $\Delta V=+300$ to
+600\,mV the TMR ratio increases and abruptly decreases for larger
$\Delta V$.  A similar almost symmetric behavior is observed for
negative $\Delta V$, except that the peak at $-600$\,mV is less
pronounced.

\begin{figure}
\includegraphics[width=7cm]{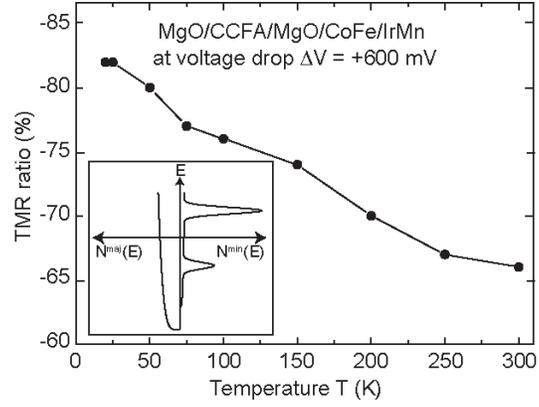}
\caption{Temperature dependence of the TMR ratio for a MgO/MgO(40
nm)/CCFA(25 nm)/MgO(3 nm)/CoFe(5 nm)/IrMn(15 nm) MTJ measured at a
voltage drop $\Delta V=+600$\,mV. Inset: Schematic spin-split DOS
of CCFA, see text.} \label{f-temp}
\end{figure}

In Fig.~\ref{f-temp} we plot the temperature dependence of the TMR
ratio measured at $\Delta V=+600$\,mV. In contrast to MTJs with
Co$_2$MnSi and Co$_2$MnGe Heusler
electrodes,\cite{hutten1,yamamoto,sakuraba} our MTJs show only a
moderate temperature dependence.  The TMR ratio increases from $-66$\%
at RT to $-84$\% upon cooling down to 20\,K. Similar behavior was
reported in Ref.~\onlinecite{yamamoto} for fully epitaxial
CCFA/MgO/CoFe MTJs.

The most striking features of our results are (i) the dominant
large \textit{inverse} TMR, (ii) its strong dependence on the
voltage drop $\Delta V$, and (iii) the weak temperature
dependence.  Thomas \textit{et al.}\cite{hutten2} reported on
inverted spin polarization of Co$_{2}$MnSi and Co$_{2}$FeSi
Heusler alloys, which gave rise to a small inverse TMR ratio of
about $-3$\% at bias voltages exceeding $-1.3$\,V. For small bias,
however, normal TMR of up to 100\% was prevailing.  Yamamoto
\textit{et al.} \cite{yamamoto} also observed a crossover to small
negative TMR at negative bias voltage in MTJ structures comprising
a Co$_2$MnGe electrode.  These authors concluded from the
conductance measurements that the inverse TMR was due to direct
tunneling and thus reflected the bulk spin-dependent electronic
density of states (DOS) of the Co$_{2}$MnGe electrode.  In our
case, the strong variation of the TMR ratio with the voltage drop,
which clearly deviates from the usually found cusp-like behavior,
and the weak temperature dependence also suggests a strong
influence of the DOS on the TMR. In the framework of
Julli\`{e}re's model, the TMR ratio can only be negative when the
effective spin polarizations P$_L$ and P$_R$ on the left and right
side of the barrier, respectively, are of opposite sign.  Sharp
features in the spin-split DOS of the electrodes give rise to a
bias dependence of the effective polarizations and thus the TMR
ratio.\cite{sharma} It was shown, see
Refs.~\onlinecite{detereza,detereza1,pettifor}, that the nature of
the bonding at the ferromagnet-insulator interface can influence
the character of the tunnelling electrons and thus both size and
sign of the effective polarization.  Due to the preferential (110)
orientation of our CCFA films, the bonding at the CCFA/MgO in our
TMR structures is significantly different from the commonly found
(100) orientation.  The related differences in the band structures
could be the reason for the inverse TMR ratios for certain $\Delta
V$ in our experiment.  Assuming that CCFA shows much sharper
features in the spin-split DOS than CoFe, \textit{e.g.} due to a
(pseudo-) gap and band edges, the data in Fig.~\ref{f-voltage} can
be qualitatively explained by the schematic spin-split DOS of CCFA
in the inset of Fig.~\ref{f-temp}.  We have also assumed that the
effective polarization of the CoFe/MgO interface is positive.  If
it is negative, minority and majority spin directions in the inset
of Fig.~\ref{f-temp} must be exchanged.  The large interval of
300-400\,mV between the Fermi level and the onsets of the peaks in
the model DOS of CCFA explains the much weaker temperature
dependence than found for other systems.\cite{sakuraba}
Theoretical studies of the spin-split DOS for CCFA/MgO(110)
interfaces and spin-dependent transport as finite voltages beyond
the Julli\`{e}re model could give more insight into the origin of
the large inverse TMR effect.  In particular, such calculations
could also properly take into account the spin and bias dependence
of the transmission probabilities, which have been shown to
strongly affect the TMR effect.\cite{heiliger}

In conclusion, we observed a large inverse TMR effect in magnetron
sputtered CCFA/MgO/CoFe MTJs at RT. The TMR ratio shows an unusual
dependence on voltage drop $\Delta V$ across the structure with large
negative values of up to $-66$\% at $\Delta V=\pm600$\,mV and small
positive values around zero bias.  The temperature dependence is
moderate with an increase from $-66$\% to $-84$\% ($\Delta
V=+600$\,mV) upon cooling from RT to 20\,K. We proposed that these
findings are related to density-of-states effects of the CCFA
electrode or the CCFA/MgO interface.  Further investigations, both
experimental and theoretical, are needed in order to understand and
control the large inverse TMR as well as its bias and temperature
characteristics.  From the application point of view, our results are
of high relevance, as they combine a large TMR ratio at a relatively
high output voltage with a moderate temperature dependence.

\newpage
%Fig1

%Fig2

%Fig3

\end{document}